\documentclass[useAMS,usenatbib]{mn2e}

\usepackage{graphicx}
\newcommand{\lya}{Ly$\alpha$}

\newcommand{\Lya}{\mbox{Ly$\alpha$}}

\newcommand{\kms}{\mbox{km s$^{-1}$}}
\newcommand{\cm}{cm$^{-2}$}
\newcommand{\civ}{\hbox{C\,{\sc iv}}}

\newcommand{\ovi}{\hbox{O\,{\sc vi}}}

\newcommand{\nciv}{\mbox{$N_{\rm \civ}$}}

\def\ltsima{$\; \buildrel < \over \sim \;$}
\def\simlt{\lower.5ex\hbox{\ltsima}}
\def\gtsima{$\; \buildrel > \over \sim \;$}
\def\simgt{\lower.5ex\hbox{\gtsima}}

\title[Galaxy-QSO absorber pair at $z = 5.719$]{A galaxy as the source of 
a \civ\ absorption system close to the epoch of reionization\thanks{
Based on observations collected at the European Organisation for Astronomical Research 
in the Southern Hemisphere, Chile [VLT program 79.A-0377]}}

\author[D\'{i}az et. al.]{C. Gonzalo D\'{i}az$^{1}$\thanks{E-mail:
gdiaz@astro.swin.edu.au}, Emma V. Ryan-Weber$^{1}$, Jeff Cooke$^{1}$, Max Pettini$^{2,3}$ 
\newauthor and Piero Madau$^{4}$\\
$^{1}$Centre for Astrophysics \& Supercomputing, Swinburne University of Technology, 
Mail H39, PO Box 218, Hawthorn, 3122 VIC, Australia\\
$^{2}$Institute of Astronomy, Madingley Road, Cambridge, CB3 0HA\\
$^{3}$Kavli Institute for Cosmology, Madingley Road, Cambridge, CB3 0HA\\
$^{4}$Department of Astronomy \& Astrophysics, University of California, Santa Cruz, CA 95064, USA}

\begin{document}

\date{Accepted. Received}

\pagerange{\pageref{firstpage}--\pageref{lastpage}} \pubyear{2010}

\maketitle

\label{firstpage}

\begin{abstract}

We find a bright ($L_{\rm UV}=2.5\,L^{\star}_{z=6}$)  Lyman alpha emitter 
(LAE) at redshift $z=5.719$ at a projected distance of $79$ physical kpc 
from a strong triply ionized carbon (\civ) absorption system 
at redshift $z=5.7238$ previously reported in the spectrum 
of the $z_{\rm em} = 6.309$ QSO SDSS J1030$+$0524. 
This is the highest redshift galaxy-absorber pair detected to-date,
supporting the idea that galaxy-wide outflows were already 
in place at the end of the epoch of reionization.
The proximity of this object makes 
it the most likely source of metals, consistent with models of outflows 
at lower redshift where significant observational evidence relates metal 
absorption systems with galaxies hosting outflows. 

In a typical outflow scenario, a wind of 200 \kms, active since the universe 
was only 0.6 Gyr old ($z \sim 8.4$), could eject metals out to 79 kpc 
at $z=5.719$.
Although the origin of metals in the intergalactic medium (IGM) is still 
under debate, our results are consistent with predictions from cosmological 
simulations which reproduce the evolution 
of the cosmic density of \civ\, from $z \sim 6$ to the present day based on 
outflow-driven enrichment of the IGM.

We also report two more \Lya\ emitters in this field, at
$z=5.973\pm 0.002$  and $z=5.676\pm 0.002$ respectively,
the former confirming the original identification by Stiavelli et al.
Our results suggest 
that the colour cut typically used to identify  $i$-dropouts ($i_{775}-z_{850}>1.3$)
misses a non-negligible fraction of blue galaxies with faint UV 
continuum at $z \simgt 5.7$.

\end{abstract}


\section{Introduction}
Medium and high resolution spectroscopy has 
revealed the presence of metals in the high redshift 
intergalactic medium (IGM) through the detection of 
absorption systems in the spectra of background 
quasi stellar objects (QSOs)
(e.g. \citealt{so1}; \citealt{pe3}; \citealt{si6}; \citealt{be6}; 
\citealt{be9}; \citealt{rw9}; \citealt{be11}; \citealt{si+11}). 
However, the enrichment source of these absorption 
systems is still poorly constrained. 
There is a debate as to whether these intergalactic metals 
originated in the recent past from nearby galaxies driving 
large-scale outflows of their interstellar media, or whether 
the IGM may have received a `blanket' enrichment at much 
earlier times, when the Universe became reionised by the 
first stars and galaxies.

\citet{pm5} favour the ``pre-galactic enrichment'' scenario, 
arguing that the cross-correlation function between Lyman 
break galaxies (LBGs) and 
\civ\ absorbers at $z\sim 3$, as well as the mean absorption 
line density, can be explained by enriched supernova (SN) 
driven winds from dwarf galaxies at $z>6$ \citep{mfr1}. 
These LBG progenitors could generate
metal bubbles  $\sim 100$\,kpc (comoving) in size within the same 
high-density regions of the IGM where LBGs would later form. 

On the other hand, the connection between star forming galaxies 
and metal absorption features is now firmly established at redshifts
$1 \simlt z \simlt  3$, strongly supporting the idea that  
galactic outflows, or `superwinds' as they are often referred to,
are one of the main sources of metal enrichment of the IGM, at least at these
redshifts \citep[][and references therein]{s10,si11}. 
Evidence for such outflows is provided by the nearly universal blueshift 
of interstellar absorption lines observed against the nuclei 
of galaxies undergoing vigorous star formation episodes
\citep[e.g.][]{he00,pe1,  sh03, s10}, with
the outflowing gas reaching velocities of 
up to $\sim 1000$\,\kms\ relative
to systemic \citep[e.g.][and reference therein]{pe02,qu09,de10}.
 In a recent study of galactic outflows at $z\sim 1.4$,
\citet{we9} found that wind velocities 
increase with mass and star formation rate 
($v_{\rm wind} \propto {\rm SFR}^{0.3}$).
A similar trend ($v_{\rm wind} \propto {\rm SFR} ^{0.35}$) 
had previosuly been shown to hold in ultra luminous infrared galaxies (ULIRGs) 
at redshift $z=0.042 - 0.16$ by \citet{ma5}; both \citet{ma5} and \citet{rvs5}
found that 70--85\% of ULIRGS at low redshifts ($z < 0.5$)
drive such energetic outflows.

The common occurrence of outflows in galaxies undergoing bursts 
of star formation and the obvious importance of this phenomenon 
for galaxy evolution and the enrichment of the IGM are some of the 
reasons why superwinds are attracting a great deal of theoretical 
interest. One of the goals of current theoretical work is to provide  
a physical explanation for the somewhat puzzling  redshift evolution 
of the comoving density of \civ\ ions in the IGM, $\Omega_{\civ}(z)$. 
Looking back in time,  $\Omega_{\civ}(z)$ appears
to be approximately constant during a  $\sim 3$\,Gyr
interval between $z \simeq 1.5$ and 5, and then drops 
by a factor of $\sim 3.5$ over a period of only $\sim 300$\,Myr
from $z = 4.7$ to 5.8 \citep{rw9,be9,si+11}.

In a series of papers \citep[][]{od6, do7, od8, odf9}, 
Oppenheimer and collaborators have explored different 
wind scenarios using cosmological hydrodynamic simulations 
which explicitly incorporate outflows from starburst galaxies
and concluded that momentum-driven winds give the best agreement
with the observations. Interestingly, in their simulations
the outflows travel to similar distances 
$R_{\rm turn}=80 \pm 20$\,physical kpc at all redshifts 
they considered. If this is the case in reality, it would imply that
at early epochs the metal enrichment by star-forming galaxies
affects larger volumes of the Universe than at later times. 
\citet{te10} have also highlighted the importance of feedback 
from galactic winds for understanding
the properties of metal absorption systems in the IGM. 
Among the models they considered,
those without galactic feedback could not reproduce the observed
evolution of the $\Omega_{\civ}(z)$, while momentum-driven wind 
simulations gave the best match with observations, in agreement with the 
proposal by \citet{od6}. Similarly,
\citet{cc10}  concluded that the dominant mechanism 
responsible for the metal enrichment of the IGM 
is star formation feedback from
the analysis of hydrodynamical simulations tuned to reproduce 
the star formation history of the universe from $z= 6$ to $z= 0$.
In these simulations, \civ\ and \ovi\ are mostly located 
in regions with temperature 
$T\geq 2 \times\,10^{4}$\,K that 
have been swept-up by the shocks produced 
by galactic-scale winds; absorption systems with column densities
$N_{\civ} = 10^{12}$--$10^{14}$\,\cm are associated 
with galaxies of luminosities $L \simlt L^\ast$. 

While it is generally agreed that galactic outflows are the 
most likely mechanism for enriching the IGM,
the observational evidence for such outflows
has so far been limited mostly to galaxies at $z \simlt 4$.
At higher redshifts, even $L^\ast$ galaxies are too faint 
to record individual spectral features -- other than the 
\Lya\ emission line -- with sufficient precision to measure
kinematic offsets between the stars and the outflowing gas
in individual galaxies. \citet{va9} did find such offsets
in \emph{stacked} spectra of $z = 5$ and 6 galaxies
selected from the Great Observatories Origins Deep Survey 
and indeed concluded that: 
``powerful, large-scale winds are common at high redshift''.
Nevertheless, it is of great interest to explore the outflow 
phenomenon in more detail at the highest redshifts, as it
holds the key not only to the initial stages of the chemical evolution
of the Universe, but also presumably to the escape of Lyman
continuum photons responsible for reionizing the IGM at $z > 6$.

If the general picture assembled from 
the above cosmological simulations is correct, 
most   \civ\ absorption systems at $z > 5$ 
should be on their first outward trajectory and should be found 
close to the parent galaxy \citep{odf9}. 
Then, a key observational test of these models
is to identify the star-forming  galaxies close to 
known \civ\ absorption systems
in the spectra of the highest redshifts QSOs.

One of the strongest \civ\ absorbers
in the high-$z$ surveys 
by \citet{rw9} and \citet{si+11}
is at $z_{\rm abs} = 5.7238$
towards the $z_{\rm em} = 6.309$ QSO SDSS J1030+0524,
with rest frame equivalent widths 
$W_0(\lambda 1548) = 0.65$\,\AA\ and $W_0(\lambda 1550) = 0.41$\,\AA\
\citep{rw9}.
The field of this QSO was imaged in the 
$i_{775}$ and $z_{850}$ bands with the 
Advanced Camera for Surveys (ACS)
on the \textit{Hubble Space Telescope (HST)}
by \citet{st5}. These authors reported a 
statistical excess of galaxies 
photometrically classified as $i$-dropouts 
(i.e. with $i_{775}-z_{850} > 1.3$),
and hence potentially located at $z \simgt 5.5$.
This result was later confirmed by \citet{ki9} 
as part of a larger study of $i$-dropouts around five QSOs at $z\sim 6$. 
In the latter work, $14\pm 4$ $i$-dropouts were found in the field of the 
QSO SDSS J1030+0524 ($\sim 11.3$\,arcmin$^2$) compared to 
$8 \pm 3$  expected from a random distribution of galaxies
at these redshifts \citep{gi4}.
In the absence of spectroscopic redshifts for all but one of the
$i$-dropouts -- a \Lya\ emitter at $z = 5.970$, both \citet{st5}
and \citet{ki9}
speculated that the excess galaxies may be part of the same overdensity 
that includes the QSO.  

However, it may also be the case that some of the candidate galaxies
are in fact associated with the foreground \civ\ absorber, 
particularly as two of the $i$-dropouts
lie within $\sim 15.5$\,arcsec of the QSO. 
This separation corresponds to a transverse distance of 
$\sim 90$\,kpc (physical) from the QSO line-of-sight at $z = 5.7$;
at $z = 2$--3 this is the impact parameter within which strong
\civ\ absorption is found to be associated with actively star-forming
galaxies \citep{s10}.
Thus, this field is an ideal location  
for testing whether galactic outflows are indeed responsible
for producing \civ\ absorbers at $z > 5.5$, 
and how the properties of such 
outflows compare with those at lower redshifts.

Motivated by these considerations, we have performed 
deep spectroscopic observations of the two galaxies 
closest to the sight-line to the QSO SDSS J1030+0524, 
and of the additional galaxy already known to be a \Lya\ emitter 
from the discovery paper by \citet{st5}.
This paper is organised as follows.
In Section 2 we describe our spectroscopic observations
as well as our reassessment of the photometry of the ACS images
obtained by \citet{st5}. We present our results in Section 3 and
discuss their implications in Section 4. We summarise our conclusions in Section 5.
Throughout this work we use AB magnitudes and assume a flat universe with 
$H_{0}=70$\,\kms\ Mpc$^{-1}$,
$\Omega_{\rm m}=0.3$ and $\Omega_{\lambda}=0.7$.

\section[]{Observations and Data Reduction}

\subsection{Spectroscopic Observations}

We obtained medium resolution spectra of  three galaxies in the field 
of the QSO SDSS J1030+0524 photometrically identified by \citet{st5}
to be at at $z>5.5$. 
Two of the galaxies were selected as being closest to the QSO sightline,
at transverse distances of $\sim 90$\,kpc
(assuming that they are at $z = 5.5$--6).
The third galaxy, J103024.08+052420.4, is the only object in the field for which
\citet{st5} were able to measure a spectroscopic redshift, $z=5.970$;
we re-observed this source for comparison purposes.

The spectra were obtained with FORS2 on UT1 of the ESO VLT, 
operated in multi-object spectroscopic mode
using $0.7''$ slits with the GRIS\_600z grism 
which covers the wavelength range from 7370 to 11000\,\AA. The
resolution is FWHM\,=\,4.9\,\AA\
(or 172.5\,\kms\ at 8500\,\AA), sampled with three pixels.
A total of 36 exposures of 900\,s each were acquired 
between 2007 April 24 and 2007 June 17;
two of these were of very low signal-to-noise ratio and were  excluded from the 
final stack of combined spectra. 
The total integration time of the final image was $30\,600\,$s. 

The observations were carried out in service mode and processed by ESO
using automatic pipelines. The data products consist 
of a sky-subtracted 2D image for each 900\,s exposure,
wavelength calibrated and corrected for bias, overscan, dark current and flat field. 
Given the challenging nature of these faint object observations,
we also processed the images independently of the ESO pipeline
(starting from the sky subtraction stage), using conventional \textsc{iraf} tasks.
Specifically,  we combined the two-dimensional slit image of each object 
with \textsc{imcombine} and we found a single emission line
centered along the spatial axis (slit direction) at the position 
where each object is expected to be. 
Then, we extracted individual exposures with \textsc{apall} and 
combined them in observational blocks of three exposures 
with \textsc{scombine} prior to the flux calibration. 
Once each block was flux calibrated  
(using the provided spectroscopic standard star spectra and 
\textsc{iraf} tasks \textsc{standard}, \textsc{sensfunc} and \textsc{calibrate}), 
we averaged all the blocks and confirmed the presence of an emission 
line uncontaminated by sky residuals in the final 
combined spectrum of each object. 

Having performed this independent check, and 
taking into consideration that the ESO pipeline was developed specifically
to reduce any systematic effect from the 
telescope and the instrument, 
we resumed the reduction process of ESO pipeline output files using \textsc{iraf}.
Given that the galaxies are very faint,
 the sky subtraction residuals
in each slit were removed 
with the task \textsc{background}, using a 
first order polynomial, before the extraction of individual spectra.
Automatic cosmic ray cleaning was carried out at this stage, 
with a few cosmic-ray affected pixels requiring manual cleaning.
Finally, after flux calibrating each spectrum, we combined the spectra of 
each galaxy in a weighted average, where the weight is the median flux value
of the 8--10 pixels centred on the \Lya\ emission line, which is the
only spectral feature detected.  
The results of our spectroscopic analysis are presented in Section 3.

\subsection{Photometry}

The existence of an overdensity of galaxies at $z > 5.5$ in the field
of the $z = 6.309$ QSO SDSS J1030+0524 first reported by \citet{st5}
has more recently been confirmed by \citet{ki9}.
This was not unexpected, given that both studies used the same 
ACS/\textit{HST} images, the same colour section criterion
and the same S/N limit.
It is therefore somewhat surprising that two out of the three galaxies 
we targeted, originally communicated to us by M. Stiavelli,
[in \citet{st5} no information was provided on individual 
galaxy positions, magnitudes and colours],
are not included in the later compilation by  \citet{ki9}.
Having confirmed spectroscopically that both objects are indeed
at $z > 5.5$ (at $z=5.676$ and $z=5.719$  respectively; see Section 3), 
we decided to remeasure ourselves the magnitudes $z_{850}$ 
and colours $i_{775}-z_{850}$ of all three galaxies considered here.
 
The data retrieved from the public \textit{HST} 
archive\footnote{http://archive.stsci.edu}
consist of  seven images in the $i_{775}$ band with a total 
exposure time of 5840\,s, and nine images in the $z_{850}$ 
band with a total exposure time of 11\,300\,s.
We ran \textsc{multidrizzle} on the reduced images ({\it flt} files) 
with a linear size of output pixels in arcseconds per pixel
of {\it pixscale} = 0.6.
The size of a pixel in units of the input pixel size was 
{\it pixfrac} = 1.0 and inverse-variance maps were used 
as weighting image ({\it final\_wht\_type} = {\it ivm}).
Distortion corrections were computed with the most 
updated distortion coefficient tables available from the 
Multimission Archive at the Space Telescope (MAST).

We used \textsc{sextractor} version 2.5.0 to calculate 
the magnitudes and colours of the three galaxies in our sample 
in order to have a consistent comparison with previous 
works in the same field. Photometry was performed in 
pseudo-dual image mode in the $z_{850}$ band and dual 
image mode in the $i_{775}$ band with the $z_{850}$ as detection image. 
The weight map from \textsc{multidrizzle} was 
used in the detection image and the corresponding 
exposure time map was used in the measurement images. 
We adopted MAG\_AUTO as the total magnitude of each source. 
Zeropoint magnitudes were obtained from the 
Space Telescope Science Institute 
website\footnote{http://www.stsci.edu/hst/acs/analysis/zeropoints} 
updated on May 19th, 2009. 
The values corresponding to F775W and F850LP filters 
for the date of observation 
are 25.67849 and 26.86663 respectively. 
Regarding extinction correction, the same values determined 
by \citet{ki9} were used: 0.048 and 0.036 for the $i_{775}$ and 
$z_{850}$ magnitudes
respectively. 
Colours give in Table~\ref{mag} were computed using 
isophotal photometry (MAG\_ISO).

Owing to correlated noise introduced by 
\textsc{multidrizzle}  in the final image, 
the errors in magnitude and flux determination
provided by \textsc{sextractor} are incorrect. 
To overcome this problem, the first step is to calculate the following
correction factor, $A$, valid in the case {\it pixscale} ($s$) $\leq$ {\it pixfrac} ($p$)
\[
A=(\frac{s}{p}-\frac{s^2}{3p^2})^2.
\]
Then, we applied the following correction to the error in the flux estimation 
and the error in the magnitude
\[
f^c_{\rm err}=\sqrt{\frac{f_{\rm err}^2}{A}-\frac{f}{g\cdot A}+\frac{f}{g}}
\]
\[
MAG^c_{\rm err}= \frac{f^c_{\rm err}}{f} \ast \frac{2.5}{\ln (10)},
\]
where $f_{\rm err}$ is the flux error estimated by \textsc{sextractor},
 $f$ is the measured flux and $g$
is the gain. 

In Table \ref{mag} we list the magnitudes and colours of the three galaxies
corrected for correlated noise and galactic extinction.
For the one galaxy in common with \citet{ki9}, J103024.08+052420.41,
we find $z_{850} = 25.94 \pm 0.05$ and
$i_{775}-z_{850} = 2.25$; for comparison \citet{ki9} reported 
$z_{850} = 25.74$ and $i_{775}-z_{850} = 2.12$ (object A13 in their Table 3).

\section{Results} 

\subsection{Lyman alpha detections}

As shown in Figs. 1, 2, and 3,
we detect a single emission line in each of the three
galaxies observed. 
We interpret these features as \Lya\ emission lines
at redshifts $z_{\rm em} = 5.7$--6.0, for the following
reasons.

As discussed in more detail below (Section~\ref{color.com}),
the ($i_{775}-z_{850}$) colours of the galaxies
together with their faint $z_{850}$ magnitudes are
indicative of high redhifts, $z > 5.4$. The main
contaminants in this colour regime are galaxies
at $z = 1$--2; in this case, the emission lines
we have detected could conceivably be
[O\,{\sc ii}]\,$\lambda\lambda 3726, 3729$.
at $z_{\rm em} = 1.2$--1.3.
However, the [O\,{\sc ii}]
doublet, with a wavelength 
separation of $2.8 \times (1+z) \simeq 6.2 $\,\AA\
should be resolved in our spectra, which have
a resolution of 4.9\,\AA\ FWHM.
Furthermore, even if the individual doublet lines
were not fully resolved, we expect the width of
the features to be FWHM\,$\simgt 8$\,\AA,
broader than FWHM\,$ \simeq 4.9$ and 6.0\,\AA\
we measure for the two newly discovered 
emission lines reported here, in Targets~2 and 3 respectively. 
The emission line in Target~1 was independently identified
as \Lya\ by \citet{st5} on the basis of its asymmetric
profile which is characteristic of resonantly scattered \Lya\
emission in galaxies undergoing large-scale outflows
\citep[e.g.][]{qu09}. We conclude that the most likely
interpretation is that all three features are
\Lya\ emission lines at redshifts $z_{\rm em} = 5.7$--6.0,
as detailed below.

We measured the values of $z_{\rm em}$ for
each galaxy from the peak of the emission line;
we estimate the uncertainty of the redshifts to be $\pm 0.002$
from consideration of the seeing and spectral resolution of our
data. However, it should be borne in mind that
the \Lya\,forest could absorb a significant fraction of the emission 
on the blue wing of the line, leading to an overestimate of the redshift.

We also measured the line flux,
from which we calculate  $L_{\rm Ly\alpha}$,
the Ly$\alpha$ luminosity
in our cosmology. This in turn can be used
to estimate the 
star formation rate, SFR$_{\rm Ly\alpha}$,
using Kennicutt's (1998) calibration of SFR$_{\rm H\alpha}$
and case B recombination

\begin{equation}
{\rm SFR}_{\rm Ly\alpha}  ({\rm M}_{\odot} {\rm yr}^{-1})
=  9.1 \times 10^{-43} L_{\rm Ly\alpha} ({\rm erg s^{-1})}.
\label{eq:SFR_Lya1}
\end{equation}

The above conversion assumes a  standard Salpeter (1955) 
stellar initial mass function (IMF);
adopting the Chabrier (2003) IMF which has proportionally 
fewer low mass stars would reduce the values of
SFR$_{\rm Ly\alpha}$ by a factor of $\sim 1.8$.
On the other hand, it is well known that the 
uncorrected \lya\ emission line luminosity can lead 
to significant \textit{under}estimates of the SFR -- not
only because of overlapping absorption by the \lya\ forest,
but especially as a consequence of resonant scattering. 
Multiply scattered \lya\ photons can be absorbed
by even small quantities of dust  and can diffuse
spatially over extended halos greater than the 
projected size of the spectrograph slit (Steidel et al. 2011).
We now briefly discuss each galaxy in turn.

\subsubsection{Target 1: J103024.08+052420.41}\label{sec.target1}

\begin{figure}
\includegraphics[width=84mm]{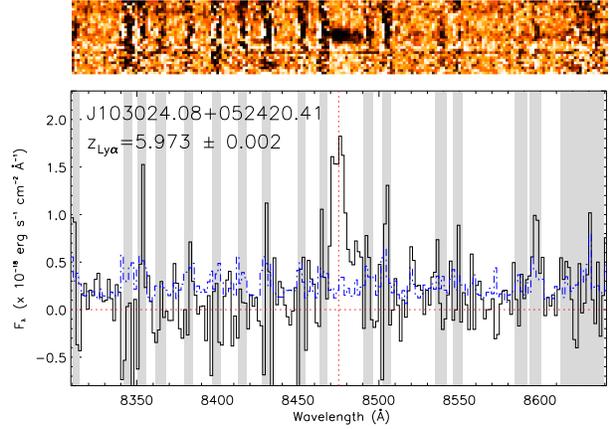}
 \caption{FORS2 spectrum of J103024.08+052420.41. 
Top: 2D image of the spectrum. The $x$-axis 
is wavelength and the $y$-axis is the spatial direction along the slit.
Bottom: 1D spectrum, with the dotted line showing the 1$\sigma$ error spectrum.
Grey regions indicate positions of all the sky lines present (with different intensities).
In both 2D and 1D spectra, an emission line is clearly 
detected at 8477\,\AA;  we identify this feature
as \Lya\ at $z = 5.973$.
}
  \label{fig:target1}
\end{figure}

This is the only galaxy  with 
spectroscopic confirmation in the sample of \citet{st5}.
We detect a clear ($\sim 15 \sigma$) emission line (see Fig. \ref{fig:target1})
which we identify as \lya\ at $z_{\rm Ly\alpha}=5.973\pm 0.002$,
in good agreement with the redshift  $z_{\rm Ly\alpha}=5.970$ reported by
\citet{st5}.
The total line flux is $F_{\rm Ly\alpha}=22 \pm 1.5\,\times 10^{-18}$\,erg s$^{-1}$ \cm, 
measured by integrating over $\sim 23$\,\AA\ ($\sim 800$\,\kms) of the spectrum;
it is not possible to compare with the data by \citet{st5}
because these authors did not report a measurement of the line flux.
The \lya\ line flux corresponds to a luminosity
$L_{\rm Ly\alpha}=8.6\pm 0.6\,\times 10^{42}$\,erg s$^{-1}$
in our cosmology, and a star formation rate
SFR$_{\rm Ly\alpha}= 7.8\pm 0.5$\,M$_{\sun}$ yr$^{-1}$ 
with the conversion (and caveats) given above (see Table \ref{mag}).
 
\subsubsection{Target 2: J103027.98+052459.51}\label{sec.target2}

\begin{figure}
\includegraphics[width=84mm]{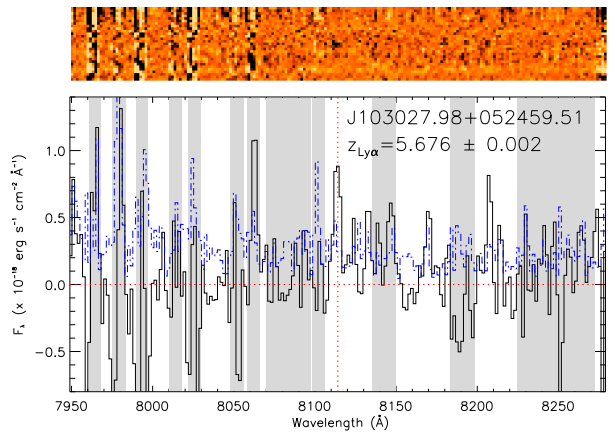}
 \caption{FORS2 spectrum of J103027.98+052459.51. Details as in Figure \ref{fig:target1}.}
  \label{fig:target2}
\end{figure}

As can be seen from Fig. \ref{fig:target2},
we detect a weak and narrow emission line which 
we interpret as \Lya\ emission at $z = 5.676\pm 0.002$. 
With a line flux $F_{\rm Ly\alpha}=4.9\pm 1.1 \times 10^{-18}$\,erg s$^{-1}$\,\cm,
the detection is significant at the $\sim 4.5 \sigma$ level. 
Although the line is weak, it does fall in a relatively clean region of
the spectrum, free from obvious residuals from the subtraction of sky emission lines
(see top panel of Fig. \ref{fig:target2}).
The corresponding luminosity and star formation rate are,
respectively,
$L_{\rm Ly\alpha}=1.7 \pm 0.4 \times 10^{42}$\,erg s$^{-1}$ 
and SFR$_{\rm Ly\alpha} = 1.6\pm 0.4$\,M$_{\sun}$ yr$^{-1}$ 
(see Table \ref{mag}).
The line has FWHM\,$\simeq  150$\,\kms\ after correcting
for the instrumental resolution.
Although this galaxy is relatively close to the QSO sight-line,
at a projected distance of 83\,kpc,
no \civ\ absorption at its redshift was
reported by either \citet{rw6} nor \citet{si+11}.

\subsubsection{Target 3: J103026.49+052505.14}\label{sec.target3}

\begin{figure}
\includegraphics[width=84mm]{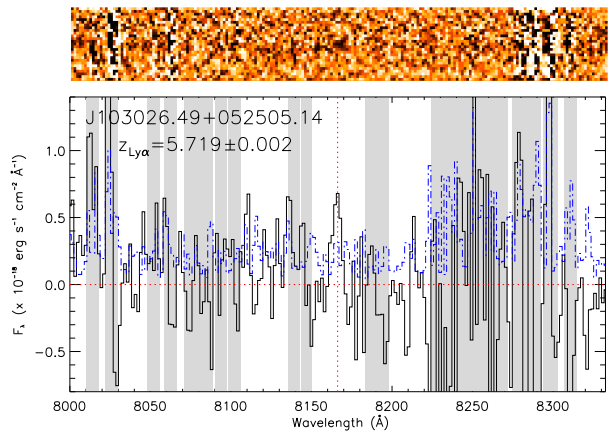}
 \caption{FORS2 spectrum of J103026.49+052505.14. Details as in Figure \ref{fig:target1}.}
  \label{fig:target3}
\end{figure}

At a projected distance of 79\,kpc, this galaxy is the closest to the QSO
sight-line among the three galaxies targeted by our observations.
We detect a weak emission line centred at $\lambda_{\rm obs} = 8168$\,\AA,
in a clean region of the spectrum, free from strong sky line residuals
(see Fig. \ref{fig:target3}).
We identify the feature as \Lya\ emission at
$z = 5.719 \pm 0.002$, with a flux
$F_{\rm Ly\alpha}=4.3 \pm 0.9 \times 10^{-18}$\,erg s$^{-1}$ \cm\
($\sim 4.8 \sigma$ level detection).
This corresponds to a luminosity
 $L_{\rm Ly\alpha}=1.5 \pm 0.3\times 10^{42}$\,erg s$^{-1}$
and star formation rate 
SFR$_{\rm Ly\alpha}= 1.4 \pm 0.3$\,M$_{\sun}$ yr$^{-1}$.

The emission line redshift differs by $\Delta v = -214$\,\kms\
from the redshift $z_{\rm abs} =  5.7238 \pm 0.0001$ of the strongest \civ\ 
absorption system in the spectrum of the QSO, with column density
\nciv\,$= 2.3\,\times 10^{14}$ \cm\ \citep{rw6}.
More recently, \citet{si+11} reported $z_{\rm abs} = 5.72438$,
corresponding to $\Delta v = -240$\,\kms,
and \nciv\,$= 3.9\,\times 10^{14}$ \cm\
for this absorber, measured
from near-IR spectra obtained with the Folded
Port Infrared Echellette (FIRE) instrument on the 6.5\,m Magellan telescope.

The velocity difference between the \Lya\ emission
line and the \civ\ system may be due to random error ($\sim 2$--$3 \sigma$)
and/or systematic errors in the wavelength scale of either the FORS2
spectrum reported here or the near-IR spectra analysed by
\citet{rw6} and \citet{si+11}. 
Alternatively, it may be an indication that the geometry
of the outflowing (or perhaps inflowing) gas
associated with this galaxy is not spherically symmetric.
The velocity offset could be larger if the \Lya\ emission line
is redshifted relatively to the systemic redshift of the galaxy,
as is normally the case in galaxies driving large scale outflows
\citep[e.g.][]{pe1,s10}.  In any case, as we discuss below, this galaxy
is sufficiently close to the QSO sight-line for an outflow at the canonical
$v_{\rm out} = 200$\,\kms\ to have travelled the projected
distance of 79\,kpc in $\sim 0.4$\,Gyr,  less than half the
age of the Universe at $z = 5.7$.

\begin{table*}
\begin{minipage}{110mm}
 \caption{Ly$\alpha$ and UV parameters, magnitudes and colours of the targets. }
 \label{mag}
 \begin{tabular}{@{}lccc}
  \hline
  \hline
  Object & Target 1 & Target 2 & Target 3 \\
  \hline
 R.A. (J2000)& 10 30 24.08 & 10 30 27.98 & 10 30 26.49 \\
 Dec. (J2000) & +05 24 20.41 & +05 24 59.51 & +05 25 05.14\\ 
 $z_{850}^a$ (mag) & $25.94\pm 0.05$ & $26.03\pm 0.06$ & $25.34\pm 0.05$\\ 
 $(i_{775}-z_{850})^b$ & $2.25$ & $1.23$ & $1.59$\\ 
  $z_{\rm Ly\alpha}^c$ &$5.973\pm 0.002$ & $5.676\pm 0.002$ & $5.719\pm 0.002$ \\
  $L_{\rm Ly\alpha}$ ($\times 10^{42}$ erg s$^{-1}$)&  $8.6\pm 0.6$ & $1.7\pm 0.4$ & $ 1.5\pm 0.3$\\ 
  SFR$_{\rm Ly\alpha}$ (M$_{\sun}$ yr$^{-1}$) &   $7.8\pm 0.5$ & $1.6\pm 0.4$ & $1.4\pm 0.3$\\ 
  $F_{\rm UV}^d$\,($\times 10^{-30}$ erg s$^{-1}$ \cm\,Hz$^{-1}$) & $1.13\pm0.08$&$1.29\pm0.08$&$2.6\pm 0.1$\\ 
  $L_{\rm UV}$ ($\times 10^{28}$ erg s$^{-1}$ Hz$^{-1}$)&  $6.4\pm 0.4$ & $6.8\pm 0.4$ & $13.6\pm 0.7$\\
  SFR$_{\rm UV}$ (M$_{\sun}$ yr$^{-1}$) &  $8.0\pm 0.5$ & $8.5\pm 0.5$ & $17.0\pm 0.9$\\ 
  Physical distance$^e$ (kpc)  & $326$ & $83$	& $79$\\
\hline

 \end{tabular}
$^{\it a}${Calculated using FLUX\_AUTO.}\\
$^{\it b}${Calculated using FLUX\_ISO.}\\
$^{\it c}${Vacuum heliocentric.}\\
$^d${Corrected for \Lya\ emission and \Lya\ forest.}\\
$^e${Transverse distance to the QSO line-of-sight.}
\end{minipage}
\end{table*}

\subsection{UV luminosities}

In this section we derive the UV continuum
fluxes of the galaxies from their $z_{850}$ magnitudes
and use them to deduce values
of the star formation rate SFR$_{\rm UV}$ for comparison
with the values of SFR$_{\rm \Lya}$ measured above.
At $z = 5.7$--6.0, the $z_{850}$ filter samples
the far-UV spectrum where, according to \citet{ma98}

\begin{equation}
{\rm L}_{\rm UV} ({\rm erg s^{-1} Hz^{-1})
= 8.0 \times 10^{-27} {\rm SFR}_{\rm UV} ({\rm M}_{\odot} {\rm yr}^{-1})} 
\label{eq:SFR_UV1}
\end{equation}

\noindent for a Salpeter IMF. 
As mentioned earlier, an IMF with fewer low-mass stars, such as 
that appropriate to the Milky Way \citep{ch3}, would lead to lower
values of SFR by a factor of $\sim 1.8$. The conversion
in eq. (\ref{eq:SFR_UV1}) applies to the UV continuum at 1500\,\AA,
but the intrinsic (i.e. before reddening) UV slope of starburst galaxies
is approximately flat in $F_\nu$ between 1500 and 1200\,\AA,
so that the above conversion should still apply. On the other hand,
the reddening correction is unknown  for our galaxies; thus the
values of SFR$_{\rm UV}$ we derive are strictly lower limits.
Furthermore, eq. (\ref{eq:SFR_UV1}) is valid for the ideal case
of a continuous star formation episode lasting for 
more than 100\,Myr; younger ages would lead to larger values
SFR$_{\rm UV}$.

Two corrections that we can apply, 
knowing the wavelengths of the \Lya\ emission lines and the transmission
curve of  the ACS F850LP filter,
are for absorption by the Ly$\alpha$ forest and for the 
contribution to the $z_{850}$ magnitudes by the \Lya\ emission line itself.
For the latter, we simply subtracted the flux of the Ly$\alpha$ line 
from the flux measured by SExtractor in the $z_{850}$ band 
(FLUX\_AUTO).
The correction for Ly$\alpha$ forest absorption is a multiplicative 
factor accounting for the fact that not all the filter 
was illuminated in wavelength space. 
We applied the following correction to amend for this effect 
\[
f^{c2}_{\nu_{0}}=f^{c}_{\nu_{0}} \cdot \frac{A_{T}}{A_{F}}\\
\]
where $f^{c}_{\nu_{0}}$ is the UV flux  corrected for \Lya\ emission,  
$A_{T}$ is the total area under the throughput curve and $A_{F}$ is the area 
under the same curve for wavelengths
$\lambda > \lambda_{\rm obs}$(\Lya) (thus assuming that, to a first approximation,
there is negligible flux below the \Lya\ emission line).

As can be seen from  
Table \ref{mag},
SFR$_{\rm UV} > {\rm SFR}_{\rm Ly\alpha}$ for all three galaxies.
As mentioned above (Section 3.1), there are many possible reasons
which can explain the common finding that the \Lya\ line luminosity
underestimates the true star formation rate in galaxies.
It is interesting that, out of the three galaxies observed, it is the one
with the clearest evidence for a large-scale outflow (Target 3) that
shows the largest difference between SFR$_{\rm UV}$ and 
$ {\rm SFR}_{\rm Ly\alpha}$.
As discussed by \citet{s11}, galaxy-scale outflows naturally lead 
to a diffuse, extended halo of \Lya\ emission which is seldom captured 
in its entirety with slit spectroscopy.

\subsection{Color confirmation}\label{color.com}

Having confirmed that all three targets are indeed at $z = 5.5$--6,
we comment briefly on the colour selection for $i$-dropouts.
According to \citet{mal5}, the colour cut $i_{775}-z_{850}>0.9$ 
in sources with $z_{850}\leq 27.5$ 
will include galaxies at $z>5.4$ with a completeness of $\sim 80$\%. 
However, galaxies at intermediate redshifts ($z \simeq  1$--2) also
have $i_{775}-z_{850} \sim 1$. 
Therefore, a color cut $i_{775}-z_{850}>1.3$ may be a stronger criterion 
for eliminating lower redshift interlopers, but it was noticed by \citet{mal5} 
that it would lead to higher incompleteness 
($\sim 20$--30\%) for $z\sim 6$ galaxies. 

Concerning our three galaxies, we note that Target 1 has 
the reddest colour, $i_{775}-z_{850}=2.25$, but the lowest UV flux, once
corrections have been applied for \Lya\ forest absorption and \Lya\ emission
(see Table \ref{mag}). The reason is that the \Lya\ emission line makes
a significant contribution to the flux measured in the $z_{850}$ filter.

With $z_{850}=26.03\pm 0.06$ Target 2 is the faintest of the three in this band,
even though its corrected UV flux is not the lowest. Note that with
$i_{775}-z_{850}=1.23$  this galaxy would be missed by a colour
selection $i_{775}-z_{850}>1.3$, even though it is intrinsically bright:
its UV luminosity, $L_{\rm UV} = 6.8 \times 10^{28}$\,erg s$^{-1}$ Hz$^{-1}$,
corresponds to $L_{\rm UV} =1.25 L^{\star}_{\rm z=6}$, adopting
$L^{\star}_{\rm z=6}=5.44 \times 10^{28}$\,erg s$^{-1}$ Hz$^{-1}$
from \citet{bo7}.

Target 3, which we associate with the \civ\ absorber at $z = 5.7238$,
has the brightest $z_{850}$ magnitude of the three objects 
($z_{850}=25.34\pm 0.05$\,mag) 
and is also the most luminous with $L_{\rm UV}=2.5\,L^{\star}_{\rm z=6}$.
In this case, the weak \Lya\ line does not make a significant contribution
to the integrated flux in the F850LP filter, resulting in a smaller correction
to the $z_{850}$ magnitude than for Target 1.

In conclusion, while there is evidence that the colour cut $i_{775}-z_{850}>1.3$ 
is effective against low redshift interlopers, our (admittedly limited) results 
show that it can also miss a significant proportion of
luminous ($L > L^\star$) galaxies at $z > 5.5$. 
Possibly, a less extreme colour cut, such as $i_{775}-z_{850}>1.2$,
may be more suitable for galaxies at $z \simgt 5.7$  with faint \Lya\ emission. 

\section{Discussion} 

\subsection{Connection between star forming galaxies and \civ\ absorption systems}

To recap, in order to investigate the origin of \civ\ absorbers at $z \sim 6$,
we have examined the field of the QSO SDSS J1030+0524, 
whose spectrum shows the
strongest \civ\ absorption doublet  in the survey by \citet{rw9},
and discovered two  \Lya\ emitters at $z \simeq 5.7$ close to the 
QSO sight-line (and confirmed a previously known \Lya\ emitter  
at $z \sim 5.97$). 
The redshift of  J103026.49+052505.14 (Target 3), which is the galaxy
closest to the QSO and the most luminous of the three,
differs by only $\Delta v  \simeq -230$\,\kms\ 
from that of the \civ\ absorption 
lines\footnote{Averaging the values of
$\Delta v$ implied by the measurements of $z_{\rm abs}$\,(\civ)
reported by \citet{rw6} and \citet{si+11}---see Section~\ref{sec.target3}.};
this finding is highly suggestive of an association between galaxy 
and  QSO absorber.

At lower redshifts ($z = 2$--3), the galaxy-QSO absorber connection has now
been quantified in some detail \citep[e.g.][]{ad3, ad5, cr10, s10}.
In the last of these studies in particular, the authors derived a rough
relationship between the equivalent width of the stronger member
of the \civ\ doublet, $W_0(\lambda 1548)$, and the projected distance
between the galaxy and the QSO sight-line \citep[see Fig. 21 of][]{s10}.
It is thus of interest to verify whether  this relationship also holds
at much higher redshifts, at least in the case of J103026.49+052505.14.

The first point to note is that,
with a SFR$_{\rm UV} = 17.0$\,M$_{\sun}$ yr$^{-1}$,
J103026.49+052505.14 is typical of the galaxies driving
outflows at redshifts $z = 2$--3 in the sample of \citet{s10}.
Second, the rest-frame equivalent width of \civ\,$\lambda 1548$ at
$z_{\rm abs} = 5.7238$ in the spectrum of SDSS J1030+0524
 is $W_0(\lambda 1548) = 0.65$\,\AA\ \citep{rw9}.
This value is intermediate
between the last two points in the $W_0(\lambda 1548)$ vs. 
galactocentric impact parameter $b$ plot constructed by \citet{s10},
between which the line equivalent width drops dramatically
from $W_0(\lambda 1548) = 1.2 \pm 0.15$\,\AA\ at $b = 63$\,kpc
to $ W_0(\lambda 1548) =  0.13 \pm 0.05$\,\AA\ at $b = 103$\,kpc. 
If we simply interpolate between these two points,
we find that $W_0(\lambda 1548) = 0.65$\,\AA\ is
\emph{on average} expected to be measured at 
impact parameters $b \simeq  70$\,kpc.
For comparison, the 13.5\,arcsec separation on the
sky between the galaxy J103026.49+052505.14 and the 
QSO SDSS J1030+0524 corresponds to an impact
parameter $b = 79$\,kpc at $z_{\rm abs} = 5.7238$.
Given the uncertainties in the interpolation and
the likely spread of the average relationship between
$W_0(\lambda 1548)$ and $b$ when applied to any 
individual galaxy, we conclude that the strength
of the \civ\ absorption seen 79\,kpc from the 
galaxy J103026.49+052505.14 
is typical of the galaxy-scale outflows operating 
at redshifts $z = 2 - 3$.

\subsection{Outflow Speed}\label{outf.speed}

With only the \Lya\ line at our disposal, 
we do not have any information on the speed of the 
outflow from J103026.49+052505.14 which has spread
metals over radii of at least 79\,kpc. Furthermore,
in the absence of the IR photometry necessary
to characterise the spectral energy distribution of this
galaxy, we have no means to estimate the age of the
starburst. Nevertheless, it is instructive to consider
some relevant timescales.

The analysis by \citet{s10}, together with
the results of high resolution spectroscopy of
gravitationally lensed star-forming galaxies
at $z = 2$--3 \citep[e.g.][and references therein]{pe02, qu09, de10},
has shown that while galactic winds may be launched 
with speed of typically $\sim 200$\,\kms,
the gas accelerates outwards reaching speeds of up
to 800\,\kms\ before reaching the virial radius.
If the bulk of the gas outflowing from J103026.49+052505.14
were indeed moving at $v_{\rm out} = 800$\,\kms,
it would reach distances of $\sim 80$\,kpc -- and could
produce absorption in the QSO spectrum -- in only
$\sim 100$\,Myr. 
 
On the other hand, other authors have considered 
more modest outflow velocities, largely on theoretical grounds. 
\citet{ma10} argued that average outflow speeds of less than
200\,\kms\ are required to match the line-of-sight
correlation function of \civ\ absorbers at redshifts $z = 1.7$--4.5.
\citet{od6} favour such velocities because in their cosmological
simulations they are found to be sufficiently high 
to enrich the low density IGM and yet low enough 
to avoid overheating it.
More recently, \citet{khm10} pointed out that 
the rapid rise in the intergalactic density of \civ\ from
$z = 5.8$ and 4.7 discovered by \citet{rw9} and \citet{be9}
may be inconsistent
with the near-complete
reionization of the IGM by $z \simeq 6$ and 
the slower build-up of stellar mass during this epoch.
All of these observations can be reconciled, however,
if there is a delay of $\sim 400$--700\,Myr between 
the onset of star formation and the appearance of its
products (i.e. the metals) at distances of $\sim 100$\,kpc
from the sites of productions. The corresponding outflow
speeds implied by this argument are also of 
order $\sim 200$\,\kms.

Adopting a more modest value
$v_{\rm out} = 200$\,\kms\
for the outflow from J103026.49+052505.14
would imply a travel time $t \sim 400$\,Myr
to reach the QSO sight-line. In turn, this
would place the onset of star formation in this
galaxy at $z = 8.4$ where galaxies are now
claimed to be found in increasing numbers
\citep[e.g.][]{bo10, bo11}. In future, 
deep observations of this and other galaxies in
the field of the QSO SDSS J1030+0524
at infrared wavelengths should allow a rough
determination of the age of the galaxy
from which an estmate of the
speed of the outflows may be inferred.

\subsection{Contribution to the IGM}

\citet{s10} showed that, given the typical extent ($\approx 90\,$kpc)
of metal-enriched regions around star-forming galaxies
at $z \sim 2$,  galaxies brighter than $\sim 0.3 L^{\star}$
can account for approximately half of the number of
\civ\ absorption systems 
with $W_0(\lambda 1548)\geq 0.15$\,\AA\
seen in QSO spectra at this average redshift.
The corollary of this statement is that a galaxy with
$L \simgt 0.3 L^{\star}$ should be found within
$\approx 90\,$kpc of half the QSO absorbers,
with the remainder presumably associated with 
fainter galaxies.

Given that the properties of the outflow from
J103026.49+052505.14 are at least superficially similar
to those of star-forming galaxies at $z = 2$--3, as 
discussed above, we can repeat the same exercise
at the higher redshifts considered here. 
Specifically, 
we estimate the number of \civ\ absorption systems 
per unit of redshift (${dN}/{dz}(\civ)_{\rm pred}$) 
expected from galaxies detected in the field of 
QSO SDSS J1030+0524 at $z\sim 5.7$, 
using the model proposed by \citet{s10}. 
Through a comparison with the observed value of ${dN}/{dz}(\civ)$, 
we draw conclusions on the properties of galaxies 
contributing to the metal enrichment 
of the IGM at $z \sim 6$ and calculate the chances 
that the $z_{\rm abs}=5.7238$ \civ\ system may be produced 
by an undetected, fainter, galaxy instead of J103026.49+052505.14.

We consider the limiting magnitude in the ACS images
to estimate an upper limit to the UV luminosity of undetected galaxies.
Adopting the $z\sim 6$ luminosity function by \citet{bo7} 
($M_{\rm UV}^{\star}=-20.24\pm 0.19$, 
$\phi^{\star}=1.4_{-0.4}^{+0.6} \times 10^{-3}$\,Mpc$^{-3}$, 
$\alpha =-1.74\pm 0.16 $) we find that the ACS images of the field
are sufficiently deep to reveal galaxies  
with luminosity  $L \simgt 0.9\, L_{z=6}^{\star}$. 
Integrating the luminosity function, 
we find that the number density of galaxies 
with $L \simgt 0.9\, L_{z=6}^{\star}$
is $n_{\rm gal} \simeq 0.287\times 10^{-3}\,h^{3}$\,Mpc$^{-3}$. 
If the model by \citet{s10} is still valid at 
$z\geq 5.5$, the co-moving cross section for absorption per galaxy is 
$\sigma_{\rm gal} \simeq 1.15$\,Mpc$^2$ for
 an impact parameter $b \simeq 90$\,kpc (physical).
At the $z_{\rm abs} = 5.7238$ of the \civ\  absorber,
 the comoving pathlength per unit of redshift is 
 ${dl}/{dz}\simeq 448$\,Mpc.
 
 Thus, we want to know the predicted number of 
 absorption systems per unit of redshift 
\[
\frac{dN}{dz}(\civ)_{\rm pred}=n_{\rm gal}\, \sigma_{\rm gal}\, \frac{dl}{dz}
\]
with $W_0(\lambda 1548)\geq 0.15$\,\AA.
With the above values for $n_{\rm gal}$,  $\sigma_{\rm gal}$ 
and ${dl}/{dz}$, we find ${dN}/{dz}(\civ)_{\rm pred} \simeq 0.15$.
This estimate is $\approx 30\%$ of the observed value 
${dN}/{dz}(\civ)_{\rm obs}=0.5 \pm 0.3$ 
at redshift $\langle z \rangle=5.75$ estimated 
by \citet{odf9}, 
from the results of \citet{rw9}, and \citet{be9} 
for absorption systems with column densities
in the range $\log$\nciv\,$=14.0$--15.0\,\cm. 
The shortfall is presumably higher when we include weaker
\civ\ systems, since the \citet{s10} limit
$W_0(\lambda 1548) \geq 0.15$
corresponds to column densities 
$\log$\nciv\,$ \geq 13.6$\,\cm\
for unsaturated absorption lines,
somewhat lower than the column density of
\civ\ for which absorption line statistics are currently available 
at these high redshifts.

We conclude that, if star-forming galaxies at $z\sim 5.7$
have enriched surrounding regions to radii of $\sim 90$\,kpc,
as is the case at $z = 2$--3,  it is necessary 
to reach at least $L_{\rm UV}\sim 0.4\, L_{z=6}^{\star}$ 
($M^{\star}_{\rm UV}\sim -19.24$, i.e. one magnitude fainter 
than the ACS images considered  in this work) 
in order to detect the full population of galaxies 
linked to absorption systems with 
$\log$\nciv\,$=14.0$--15.0\,\cm.

\section{Conclusions} 

From deep spectroscopy of galaxies in the field of 
$z_{\rm em} = 6.309$ QSO SDSS J1030+0524 we have identified
three \Lya\ emitters at $z \geq 5.7$. Of these three galaxies,
the brightest ($L_{\rm UV}=2.5\,L^{\star}_{\rm z=6}$)
and closest to the QSO sight-line (impact parameter $b = 79$\,kpc),
J103026.49+052505.14,
is at a redshift $z = 5.719$ which differs by only 
$\Delta v \simeq -230$\,\kms\
from the absorption redshift of a strong \civ\ system
in the spectrum of the QSO.

The chances of a random association between
\Lya\ emitter and \civ\ absorber 
are very low.
For example, if we first consider the idealised case
of a random distribution of galaxies with 
$n_{\rm gal} \simeq 1.23\times 10^{-3}$\,Mpc$^{-3}$,
appropriate to galaxies of luminosity 
$L\simgt 0.4\, L_{z=6}^{\star}$, then there is
only a $1.1 \times 10^{-3}$ ($\sim 0.1$\%)
chance of finding such a galaxy within
a sphere of radius 90\,kpc (physical)
centred on the QSO sight-line at $z = 5.7238$.
However, for this calculation it is more appropriate to 
consider a cylinder, rather than a sphere, with radius
90\,kpc but depth along the line of sight corresponding
to $\Delta v \simeq \pm 230$\,\kms,
the velocity difference between the \Lya\ emitter
and the \civ\ absorber.  
The volume of this configuration
is approximately six times larger than that of a sphere of radius
90\,kpc, raising the probability of finding a galaxy by
chance within this volume to $\sim 0.7$\%.

A second factor to take into account is galaxy clustering.
Indeed, \citet{st5} and \citet{ki9} reported an overdensity
by a factor of $\sim 2$ at  $z > 5.5$ in the field of the QSO
J1030+0524. If this average factor applied to the environment
of the galaxy J103026.49+052505.14, the above probability
would be increased by a factor of two. More important,
however, is the possibility of significant clustering with
lower luminosity dwarf galaxies which would remain 
unrecognised given the current limits of the photometry
in this field. Extending the above calculations further down
the luminosity function, to galaxies as faint as
$\sim 0.06\, L_{z=6}^{\star}$, would increase the above 
probability by one order of magnitude.
Indeed, the $\sim 230$\,\kms\ velocity difference
between \Lya\ emission and \civ\ absorption 
could be taken as evidence for an origin
of the metals in a dwarf galaxy
in the same overdensity as  J103026.49+052505.14.
On the other hand, at present we have no way of addressing empirically
the question of whether such low-luminosity galaxies support 
large-scale outflows of metal-enriched gas,
and how their mass-outflow rates compare 
with those of the more luminous galaxies
in which the outflows are seen directly.

The result reported here is the first direct observational evidence that 
the circum-galactic medium of early luminous star-forming galaxies
was already enriched with heavy elements at $z \sim 6$. 
Even at these high redshifts, some of the properties
of the galaxy-wide outflows responsible for enriching
the circum-galactic medium appear to be superficially
similar to those of their counterparts at $z = 2$--3, which 
have been characterised in some detail. In particular,
the strength of the \civ\ absorption associated with
J103026.49+052505.14 at a radial distance of 79\,kpc 
fits in well with expectations based on data
appropriate to lower redshifts
galaxies, even though the average density of the IGM
was much higher at $z \sim 6$ than at $z = 2$--3.

We emphasise the importance of extending the available
photometry of J103026.49+052505.14 to include near-IR bands
from which the SED of the galaxy can be determined with the
accuracy required to deduce its age. With this information
in hand, it should be possible to estimate the average speed
of the outflow which has transported the metals out to
at least $\sim 80$\,kpc from the stars that produced
them. If a speed $v_{\rm out} \simeq 200$\,\kms\
is appropriate, as argued by some, the starburst episode
must have lasted at least $\sim 400$\,Myr,
placing its origin at $z \simeq 8.4$;  we now
know this to have been an epoch when star-formation in galaxies
was already underway. Higher outflow speeds would imply
later onsets of the starburst.

The observations presented here show that it is feasible,
with dedication, to identify the galaxies linked to 
QSO absorption systems even at the highest redshifts where
intervening metals have been detected in QSO spectra,
and thus provide an incentive to pursue similar searches
in other fields. Finally, we also point out that
the colour cut $i_{775}-z_{850}>1.3$ which has been
proposed to avoid low redshift interlopers can miss
a significant proportion of bluer galaxies at $z \simgt 5.7$
thus affecting mainly the LAE population.
We are pursuing this topic in a forthcoming paper.

\section*{Acknowledgments}
 
We are grateful to Massimo Stiavelli who kindly shared his ACS images
of the QSO field in advance of publication, and to Ben Oppenhaimer
and Masami Ouchi for stimulating discussions.
We thank the referee, Neil Crighton, for valuable suggestions
that improved the paper.
CGD acknowledges financial support by the Victorian State Government
through the International Research Scholarship Program;
ERW's research is supported by Australian Research Council grant DP1095600;
PM acknowledges support by the NSF through
grant AST-0908910.\\

\label{lastpage}


\begin{thebibliography}{99}
 
\bibitem[\protect\citeauthoryear{Adelberger et al.}{2003}]{ad3} Adelberger, K. L., Steidel, C. C., Shapley, A. E., \& Pettini, M. 2003, ApJ, 584, 45
\bibitem[\protect\citeauthoryear{Adelberger et al.}{2005}]{ad5} Adelberger, K. L., Shapley, A. E., Steidel, C. C., Pettini, M., Erb, D. K. \& Reddy, N. A. 2005, ApJ, 629, 636
\bibitem[\protect\citeauthoryear{Becker et al.}{2006}]{be6} Becker, G. D., Sargent, W. L. W.,  Rauch, M. \& Simcoe, R. 2006, ApJ, 640, 69
\bibitem[\protect\citeauthoryear{Becker, Rauch \& Sargent}{2009}]{be9} Becker, G. D., Rauch, M. \& Sargent, W. L. W. 2009, ApJ, 698, 1010
\bibitem[\protect\citeauthoryear{Becker et al.}{2011}]{be11} Becker, G. D., Sargent, 
W. L. W., Rauch, M., \& Calverley, A. P.\ 2011, ApJ, submitted (arXiv:1101.4399) 
\bibitem[\protect\citeauthoryear{Bouwens et al.}{2007}]{bo7} Bouwens, R. J., Illingworth, G. D., Franx, M., Ford, H. 2007, ApJ, 670, 928
\bibitem[\protect\citeauthoryear{Bouwens et al.}{2010}]{bo10} Bouwens, R. J., et al.\ 2010, ApJ, 709, L133 
\bibitem[\protect\citeauthoryear{Bouwens et al.}{2011}]{bo11} Bouwens, R. J., et al.\ 2011, ApJ, in press (arXiv:1006.4360)
\bibitem[\protect\citeauthoryear{Cen \& Chisari}{2010}]{cc10} Cen, R., Chisari, N. E. 2011, ApJ, 731, 11 
\bibitem[\protect\citeauthoryear{Chabrier            }{2003}]{ch3} Chabrier, G.\ 2003, PASP, 115, 763 
\bibitem[\protect\citeauthoryear{Crighton et al.}{2010}]{cr10} Crighton, N. H. M., et 
al.\ 2010, MNRAS, in press (arXiv:1006.4385). 
\bibitem[\protect\citeauthoryear{Dav\'e \& Oppenheimer}{2007}]{do7} Dav\'e, R. \& Oppenheimer, B. D. 2007, MNRAS, 374, 427
\bibitem[\protect\citeauthoryear{Dessauges-Zavadsky et al.}{2010}]{de10} Dessauges-Zavadsky, M., D'Odorico, S., Schaerer, D., Modigliani, A., Tapken, C., \& Vernet, J.\ 2010, A\&A, 510, A26 
\bibitem[\protect\citeauthoryear{Giavalisco et al.}{2004}]{gi4} Giavalisco, M., et al. 2004, ApJ, 600, L93
\bibitem[\protect\citeauthoryear{Heckman et al.}{2000}]{he00} Heckman, T. M., 
Lehnert, M. D., Strickland, D. K., \& Armus, L.\ 2000, ApJS, 129, 493 
\bibitem[\protect\citeauthoryear{Kennicutt}{1998}]{ken98} Kennicutt, R. C., Jr.\ 1998, ARAA, 36, 189 
\bibitem[\protect\citeauthoryear{Kim et al.}{2009}]{ki9} Kim, S., et al. 2009, ApJ, 695, 809
\bibitem[\protect\citeauthoryear{Kramer, Haiman \& Madau}{2010}]{khm10} Kramer, R., Haiman, Z. \& Madau, P. 2010 astro--ph/arXiv:1007.3581, 
submitted to MNRAS
\bibitem[\protect\citeauthoryear{Madau, Pozzetti \& Dickinson}{1998}]{ma98}	Madau, P., Pozzetti, L. \& Dickinson, M. 1998, ApJ, 498, 106
\bibitem[\protect\citeauthoryear{Madau, Ferrara \& Rees}{2001}]{mfr1} Madau, P., Ferrara, A. \& Rees M. J. 2001, ApJ, 555, 92
\bibitem[\protect\citeauthoryear{Malhotra et al.}{2005}]{mal5}	Malhotra, S., et al. 2005, ApJ, 626, 666
\bibitem[\protect\citeauthoryear{Martin}{2005}]{ma5} Martin, Crystal L. 2005, ApJ, 621, 227
\bibitem[\protect\citeauthoryear{Martin et al.}{2010}]{ma10} Martin, C., Scannapieco, E., Ellison, S., Hennawi, J., Djorgovski, S. \& Fournier, A. 2010,
ApJ, 721, 174
\bibitem[\protect\citeauthoryear{Oppenheimer \& Dav\'e}{2006}]{od6} Oppenheimer, B. D. \& Dav\'e, R. 2006, MNRAS, 373, 1265
\bibitem[\protect\citeauthoryear{Oppenheimer, Dav\'e \& Finlator}{2009}]{odf9} Oppenheimer, B. D., Dav\'e, R. \& Finlator, K. 2009, 
MNRAS, 396, 729
\bibitem[\protect\citeauthoryear{Oppenheimer \& Dav\'e}{2008}]{od8} Oppenheimer, B. D. \& Dav\'e, R. 2008, MNRAS, 387, 577
\bibitem[\protect\citeauthoryear{Oppenheimer et al.}{2011}]{opp11} Oppenheimer, B. D. \& Dav\'e, R. 2008, MNRAS, in press
\bibitem[\protect\citeauthoryear{Pettini et al.}{2001}]{pe1} Pettini, M., Shapley, A. E., Steidel, C. C., Cuby, J.-G., Dickinson, M., Moorwood, A. F. M., Adelberger, K. L., \& Giavalisco, M.\ 2001, ApJ, 554, 981 
\bibitem[\protect\citeauthoryear{Pettini et al.}{2002}]{pe02} Pettini, M., Rix, 
S. A., Steidel, C. C., Adelberger, K. L., Hunt, M. P., \& Shapley, A. E.\ 2002, ApJ, 569, 742 
\bibitem[\protect\citeauthoryear{Pettini et al.}{2003}]{pe3} Pettini, M., Madau, P., Bolte, M., Prochaska, J. X., Ellison, S. L. \& Fan, X. 2003, ApJ, 594, 695
\bibitem[\protect\citeauthoryear{Porciani \& Madau}{2005}]{pm5} Porciani, C. \& Madau, P. 2005, ApJ, 625, L43
\bibitem[\protect\citeauthoryear{Powell, Slyz \& Devriendt}{2010}]{psd10} Powell, L., Slyz, A. \& Devriendt, J. 2010 astro--ph/arXiv:1012.2839, submitted to MNRAS
\bibitem[\protect\citeauthoryear{Quider et al.}{2009}]{qu09} Quider, A. M., Pettini, 
M., Shapley, A. E., \& Steidel, C. C.\ 2009, MNRAS, 398, 1263 
\bibitem[\protect\citeauthoryear{Rupke, Veilleux \& Sanders}{2005}]{rvs5} Rupke, D. S., Veilleux, S. \& Sanders, D. B. 2005, ApJS., 160, 115
\bibitem[\protect\citeauthoryear{Ryan-Weber, Pettini \& Madau}{2006}]{rw6} Ryan-Weber, E. V., Pettini, M. \& Madau, P. 2006, MNRAS, 371, 78
\bibitem[\protect\citeauthoryear{Ryan-Weber et al.}{2009}]{rw9} Ryan-Weber, E. V., Pettini, M., Madau, P. \& Zych, B. J. 2009, MNRAS, 395, 1476
\bibitem[Salpeter(1955)]{1955ApJ...121..161S} Salpeter, E. E.\ 1955, ApJ, 121, 161 
\bibitem[\protect\citeauthoryear{Shapley et al.}{2003}]{sh03} Shapley, A. E., 
Steidel, C. C., Pettini, M., \& Adelberger, K. L.\ 2003, ApJ, 588, 65 
\bibitem[\protect\citeauthoryear{Simcoe}{2006}]{si6} Simcoe R. A. 2006, ApJ, 653, 977
\bibitem[\protect\citeauthoryear{Simcoe}{2011}]{si11} Simcoe R. A. 2011, ApJ, in press\bibitem[\protect\citeauthoryear{Simcoe et al.}{2011}]{si+11} Simcoe R. A., et al.  2011, ApJ, in press
\bibitem[\protect\citeauthoryear{Songaila}{2001}]{so1} Songaila, A. 2001, ApJ, 561, L153
\bibitem[\protect\citeauthoryear{Steidel et al.}{2005}]{s5} Steidel, C. C., Adelberger, K. L., Shapley, A. E., Erb, D. K., Reddy, N. A. \& Pettini, M. 2005, ApJ, 626, 44
\bibitem[\protect\citeauthoryear{Steidel et al.}{2010}]{s10} Steidel, C. C., Erb, D. K., Shapley, A. E., Pettini, M., Reddy, N., 
 Bogosavljevi\'c, M., Rudie, G. C. \& Rakic, O. 2010, ApJ, 717, 289
\bibitem[\protect\citeauthoryear{Steidel et al.}{2011}]{s11} Steidel, C. C., 
Bogosavljevi{\'c}, M., Shapley, A. E., Kollmeier, J. A., Reddy, N. A., Erb, 
D. K., \& Pettini, M.\ 2011, ApJ, in press (arXiv:1101.2204) 
\bibitem[\protect\citeauthoryear{Stiavelli et al.}{2005}]{st5} Stiavelli, M., et al. 2005, ApJ, 622, L1
\bibitem[\protect\citeauthoryear{Tescari et al.}{2010}]{te10} Tescari, E., Viel, M., D'Odorico, V., Cristiani, S., Calura, F., Borgani, S., Tornatore, L. 2011, MNRAS, 411, 826
\bibitem[\protect\citeauthoryear{Vanzella et al.}{2009}]{va9} Vanzella, E., et al.\ 
2009, ApJ, 695, 1163 
\bibitem[\protect\citeauthoryear{Weiner et al.}{2009}]{we9} Weiner, B. J., et al. 2009, ApJ, 692, 187


\end{thebibliography}
\end{document}